\documentclass[preprint,amsmath,amssymb,aps,showkeys,showpacs]{revtex4}
\usepackage[english]{babel}
\usepackage{graphicx}
\usepackage{graphics}
\usepackage{amsmath}
\usepackage{dcolumn}
\usepackage{amssymb}
\usepackage{bm}

\begin{document}
\title{Some aspects of an induced electric dipole moment in rotating and nonrotating frames}
\author{Abinael B. Oliveira}
\affiliation{Departamento de F\'isica, Universidade Federal da Para\'iba, Caixa Postal 5008, 58051-900, Jo\~ao Pessoa-PB, Brazil.}

\author{Knut Bakke}
\email{kbakke@fisica.ufpb.br}
\affiliation{Departamento de F\'isica, Universidade Federal da Para\'iba, Caixa Postal 5008, 58051-900, Jo\~ao Pessoa-PB, Brazil.}

\begin{abstract}

Quantum effects on a neutral particle (atom or molecule) with an induced electric dipole moment are investigated when it is subject to the Kratzer potential and a scalar potential proportional to the radial distance. In addition, this neutral is placed in a region with electric and magnetic fields. This system is analysed in both nonrotating and rotating reference frames. Then, it is shown that bound state solutions to the Schr\"odinger equation can be achieved and, in the search for polynomial solutions to the radial wave function, a restriction on the values of the cyclotron frequency is analysed in both references frames.

\end{abstract}

\keywords{rotating effects, induced electric dipole moment, Landau quantization, Kratzer potential, linear scalar potential}
\pacs{03.65.Ge, 31.30.jc, 31.30.J-, 03.65.Vf}

\maketitle

\section{Introduction}

In the literature, crossed electric and magnetic fields have attracted interests in studies of the hydrogen atom \cite{cross1,cross2,cross3,cross4,cross5,cross6}, large electric dipole moments \cite{cross7}, atoms and molecules in strong magnetic field \cite{cross8,cross9,cross10}, geometric quantum phases \cite{whw,whw2} and the quasi-Landau behaviour in atomic systems \cite{cross11,cross12}. In Ref. \cite{lin3}, it is shown that crossed electric and magnetic fields can give rise to an analogue of the Landau quantization \cite{landau} for an atom with an induced electric dipole moment. Recently, this interaction of the induced electric dipole moment of an atom with a uniform effective magnetic field produced by crossed electric and magnetic fields has been investigated in quantum rings \cite{dantas}, subject to scalar potentials \cite{ob,ob2} and in rotating reference frames \cite{dantas1,ob3}. Other interesting effects that arise from quantum fluctuations in the QED vacuum have been reported in Refs. \cite{extra,extra2}. The focus of this work is on the quantum effects on an atom (molecule) with an induced electric dipole moment subject to the Kratzer potential \cite{kratzer,kratzer2,kratzer3,kratzer6,kratzer5} and a scalar potential proportional to the radial distance in a region with crossed electric and magnetic fields. We deal with this system in both nonrotating and rotating frames. Particular interests in rotating systems arise from studies of geometric quantum phases \cite{sag,sag2,sag5,r4,ac2,cond3,cond3a,r13,anan,r14,r15}, spintronics \cite{spint1,spint2,spint3}, quantum rings \cite{r11,r12} scalar bosons \cite{castro}, DKP equation \cite{dkp} and electroweak interactions \cite{electro}. Therefore, the present study fills a lack in the studies of neutral particles with no permanent electric dipole moment that interacts with external fields.

The structure of this paper is: in section II, we introduce the quantum description of a moving neutral particle (molecule or atom) with an induced electric dipole moment in a region with electric and magnetic fields; thus, we consider the field configuration proposed in Ref. \cite{lin3} that gives rise to the analogue of the Landau quantization and analyse this system subject to the Kratzer potential \cite{kratzer,kratzer2,kratzer3} and a scalar potential proportional to the radial distance; in section III, we investigate the quantum effects on the system described in the previous section by considering a rotating reference frame; in section IV, we present our conclusions.

\section{nonrotating frame}

Let us begin this section by reviewing the quantum description of an atom or molecule with an induced electric dipole moment that moves with a velocity $v\ll c$ and interacts with external fields. As shown in Ref. \cite{whw,lin3,ob,ob2,ob3}, the Hamiltonian operator of the system is given by (with $\hbar=c=1$)
\begin{eqnarray}
\mathbb{H}_{0}=\frac{1}{2m}\left(\hat{p}+\alpha\,\vec{E}\times\vec{B}\right)^{2}-\frac{\alpha}{2}\,E^{2}+\mathbb{V},
\label{1.1}
\end{eqnarray}
where $m$ is the mass of the particle, $\alpha$ is is the dielectric polarizability, $\mathbb{V}$ is the potential energy, $\hat{p}=-i\vec{\nabla}$ is the momentum operator (vector operator), and the vectors $\vec{E}$ and $\vec{B}$ are the electric and magnetic fields in the laboratory frame, respectively. According to Ref. \cite{whw2}, the term $\alpha\,E^{2}$ given in Eq. (\ref{1.1}) is very small compared with the kinetic energy of the atoms, therefore we can neglect it without loss of generality from now on. 

The focus of this section is on the quantum effects on an atom (molecule) with an induced electric dipole moment in a region with a uniform effective magnetic field, which is subject to the Kratzer potential \cite{kratzer,kratzer2,kratzer3} and a scalar potential proportional to the radial distance. In particular, this uniform effective magnetic field is defined as $\vec{B}_{\mathrm{eff}}=\vec{\nabla}\times\left(\vec{E}\times\vec{B}\right)$, where the magnetic and electric fields in the laboratory frame are \cite{lin3}
\begin{eqnarray}
B_{z}=B_{0};\,\,\,E_{r}=\frac{\lambda\,r}{2},
\label{1.2}
\end{eqnarray}
where $B_{0}$ is a constant and $\lambda$ is a constant related to the uniform volume charge density. As observed in Ref. \cite{lin3}, the interaction of the induced electric dipole moment of the atom with the magnetic and electric fields given in Eq. (\ref{1.2}) gives rise to a discrete spectrum of energy which is known as an analogue of the Landau quantization \cite{landau}. Recently, quantum effects on this Landau-type system have been investigated in a quantum ring \cite{dantas}, under the influence of a Coulomb-type potential \cite{ob}, a linear confining potential \cite{ob2} and the Kratzer potential in a rotating frame \cite{ob3}. Besides, in order to investigate the influence of the Kratzer potential \cite{kratzer,kratzer2,kratzer3} and a scalar potential proportional to the radial distance, then, we write the potential energy $\mathbb{V}$ as follows:
\begin{eqnarray}
\mathbb{V}=b\,r-\frac{2D\,a}{r}+\frac{D\,a^{2}}{r^{2}}.
\label{1.3}
\end{eqnarray}
The first term in Eq. (\ref{1.3}) corresponds to the the scalar potential proportional to the radial distance, where $b$ is a constant. The second term corresponds to the Kratzer potential, where $D$ and $a$ are constants. It has attracted a great interest in studies of molecules \cite{kratzer4,molecule,ct5}. Thereby, after substituting Eqs. (\ref{1.2}) and (\ref{1.3}) into Eq. (\ref{1.1}), the time-independent Schr\"odinger equation becomes 
\begin{eqnarray}
\mathcal{E}\Psi&=&-\frac{1}{2m}\nabla^{2}\Psi+i\,\frac{\alpha\,\lambda\,B_{0}}{2m}\frac{\partial\Psi}{\partial\varphi}+\frac{\alpha^{2}\lambda^{2}B_{0}^{2}}{8m}\,r^{2}\Psi-\frac{2\,D\,a}{r}\,\Psi+\frac{D\,a^{2}}{r^{2}}\,\Psi+b\,r\,\Psi,
\label{1.5}
\end{eqnarray}
where $\nabla^{2}$ is the Laplacian in cylindrical coordinates. Let us take $\Psi\left(r,\,\varphi,\,z\right)=\Phi\left(\varphi\right)\,Z\left(z\right)\,F\left(r\right)$; thus, we have that $\Phi\left(\varphi\right)=e^{i\nu\varphi}$, where $\nu=0,\pm1,\pm2,\ldots$ and $Z\left(z\right)=e^{ik\,z}$, where $k$ is a constant. Then, by defining the parameter $\omega=\frac{\alpha\,\lambda\,B_{0}}{m}$ as the cyclotron frequency associated with the Landau quantization for an atom with an induced electric dipole moment \cite{lin3}, let us substitute $\Psi\left(r,\,\varphi,\,z\right)$ into Eq. (\ref{1.5}) and perform a change of variables given by
\begin{eqnarray}
y=\sqrt{\frac{m\,\omega}{2}}\,r.
\label{1.6}
\end{eqnarray} 
In this way, we obtain the following second order differential equation:
\begin{eqnarray}
F''+\frac{1}{y}\,F'-\frac{\tau^{2}}{y^{2}}\,F-y^{2}\,F+\frac{\mu}{y}\,F-\theta\,y\,F+\chi\,F=0,
\label{1.7}
\end{eqnarray}
where $\tau^{2}=\nu^{2}+2m\,D\,a^{2}$, $\mu=\frac{4m\,D\,a}{\sqrt{\frac{m\,\omega}{2}}}$, $\theta=\frac{2m\,b}{\left(\frac{m\,\omega}{2}\right)^{3/2}}$ and $\chi=\frac{2}{m\,\omega}\left[2m\mathcal{E}-k^{2}+m\,\omega\,\nu\right]$. It is well-known that the analysis of the asymptotic behaviour of Eq. (\ref{1.7}) determines the form of the function $F\left(y\right)$, therefore, this function can be written in terms of an unknown function $H\left(y\right)$ as 
\begin{eqnarray}
F\left(y\right)=e^{-\frac{y^{2}}{2}-\frac{\theta\,y}{2}}\,y^{\left|\tau\right|}\,H\left(y\right).
\label{1.9}
\end{eqnarray}
By substituting Eq. (\ref{1.9}) into Eq. (\ref{1.7}), we obtain that $H\left(y\right)$ is the solution to the following equation:
\begin{eqnarray}
H''+\left[\frac{2\left|\tau\right|+1}{y}-\theta-2y\right]H'+\left[\chi+\frac{\theta^{2}}{4}-2\left|\tau\right|-2+\frac{2\mu-\theta\left(2\left|\tau\right|+1\right)}{2y}\right]H=0,
\label{1.10}
\end{eqnarray}
which is the biconfluent Heun equation \cite{heun}, then, $H\left(y\right)=H_{\mathrm{B}}\left(2\left|\tau\right|,\,\theta,\,\chi+\frac{\theta^{2}}{4},\,-2\mu;\,y\right)$ is the biconfluent Heun function.

We wish the function $F\left(y\right)$ goes to zero when $y\rightarrow\infty$ and $y\rightarrow0$, therefore, let us use the Frobenius method \cite{arf}. In this method, we first write the biconfluent Heun function as a power series around the origin, $H\left(y\right)=\sum_{i=0}^{\infty}b_{i}\,y^{i}$, and then, we search for polynomial solutions to the biconfluent Heun equation (\ref{1.10}). By substituting this series in Eq. (\ref{1.10}) we obtain the relation
\begin{eqnarray}
b_{1}=\left[\theta-\frac{2\mu}{2\left|\tau\right|+1}\right]\,b_{0},
\label{1.11}
\end{eqnarray}
and the recurrence relation:
\begin{eqnarray}
b_{i+2}=\frac{\theta\left(i+1\right)-2\mu+\theta\left(2\left|\tau\right|+1\right)}{\left(i+2\right)\left(i+2+2\left|\tau\right|\right)}\,b_{i+1}+\frac{2i-\chi-\frac{\theta^{2}}{4}+2+2\left|\tau\right|}{\left(i+2\right)\left(i+2+2\left|\tau\right|\right)}\,b_{i}.
\label{1.12}
\end{eqnarray}

From the recurrence relation above we have that the biconfluent Heun series becomes a polynomial of degree $n$ by imposing that \cite{heun}
\begin{eqnarray}
\chi+\frac{\theta^{2}}{4}-2\left|\tau\right|-2=2n;\,\,\,\,\,\,\,b_{n+1}=0,
\label{1.13}
\end{eqnarray}
for $n=1,2,3,\ldots$. Hence, these two conditions must be analysed in order to achieve a polynomial solution. From the condition $\chi+\frac{\theta^{2}}{4}-2\left|\tau\right|-2=2n$, we obtain 
\begin{eqnarray}
\mathcal{E}_{n,\,\nu}=\frac{1}{2}\,\omega\left[n+\left|\tau\right|-\nu+1\right]-\frac{2\,b^{2}}{m\,\omega^{2}}+\frac{k^{2}}{2m},
\label{1.14}
\end{eqnarray}
which yields the energy levels of a neutral particle with an induced electric dipole moment in a region with crossed electric and magnetic fields under the influence of scalar potentials. Observe that the field configuration of electric and magnetic field given in Eq. (\ref{1.2}) gives rise to a uniform effective magnetic field in the $z$-direction that yields an analogue of the Landau quantization as pointed out in Ref. \cite{lin3}. By comparing the energy levels (\ref{1.14}) with the analogue of the Landau levels in Ref. \cite{lin3}, we have that the presence of the Kratzer potential and the scalar potential proportional to the radial distance modifies the energy levels and breaks the degeneracy of the Landau-type levels.

On the other hand, our search for polynomial solutions to the biconfluent Heun series will be completed when we analyse the condition $b_{n+1}=0$ given in Eq. (\ref{1.13}). For this purpose, let us consider $b_{0}=1$. Then, let us construct a polynomial of first degree. For $n=1$, we obtain $b_{2}=0$ from $b_{n+1}=0$. In this way, we obtain
\begin{eqnarray}
\omega_{1,\,\nu}^{3}-\frac{64m\,D^{2}a^{2}}{2\left|\tau\right|+1}\,\omega_{1,\,\nu}^{2}+\frac{32b\,Da\left(4\left|\tau\right|+3\right)}{2\left|\tau\right|+1}\,\omega_{1,\,\nu}-\frac{32\,b^{2}\left(\left|\tau\right|+1\right)}{m}=0,
\label{1.16}
\end{eqnarray}
which is a third degree algebraic equation. It means that, in order to achieve a polynomial of first degree to $H\left(y\right)$, we have that not all values of the cyclotron frequency are permitted, but only which are determined by the third degree algebraic equation (\ref{1.16}) \cite{fb4}. Despite having at least one real solution to the third degree algebraic equation (\ref{1.16}), we do not write this solution because it is very long. Besides, for each energy level $n$ of the system, we can have a different expression that determines the possible values of the cyclotron frequency. For this reason, we have labelled $\omega=\omega_{n,\,l}$ in Eq. (\ref{1.16}) and rewrite the energy levels (\ref{1.14}) in the form: 
\begin{eqnarray}
\mathcal{E}_{n,\,\nu}=\frac{1}{2}\,\omega_{n,\,\nu}\left[n+\left|\tau\right|-\nu+1\right]-\frac{2\,b^{2}}{m\,\omega^{2}_{n,\,\nu}}+\frac{k^{2}}{2m},
\label{1.17}
\end{eqnarray}
which are the energy levels of a neutral particle with an induced electric dipole moment in a region with a uniform effective magnetic field, where there exists the influence of the Kratzer potential and a scalar potential proportional to the radial distance.

\section{rotating frame}

In this section, we consider a rotating frame where the system discussed in the previous section is in a reference frame that rotates with a constant angular velocity $\vec{\Omega}=\Omega\,\hat{z}$. By following Refs. \cite{landau3,landau4,dantas,anan,r13,fb5,fb6}, the time-independent Schr\"odinger equation in the rotating frame is given by
\begin{eqnarray}
\mathbb{H}_{0}\,\Psi-\vec{\Omega}\cdot\hat{L}\,\Psi=\mathcal{E}\Psi,
\label{2.1}
\end{eqnarray}
where $\mathbb{H}_{0}$ is given in Eq. (\ref{1.1}), i.e., it is to the Hamiltonian operator in the absence of rotation, and $\hat{L}$ is the angular momentum operator. In the present case, the angular momentum operator is given by $\hat{L}=\vec{r}\times\left(\hat{p}+\alpha\,\vec{E}\times\vec{B}\right)$, where $\vec{r}=r\,\hat{r}$ in a two-dimensional system. Thereby, the time-independent Schr\"odinger equation (\ref{2.1}) becomes 
\begin{eqnarray}
\mathcal{E}\Psi&=&-\frac{1}{2m}\nabla^{2}\Psi+i\,\frac{\alpha\,\lambda\,B_{0}}{2m}\frac{\partial\Psi}{\partial\varphi}+\frac{\alpha^{2}\lambda^{2}B_{0}^{2}}{8m}\,r^{2}\Psi\nonumber\\
&+&i\Omega\,\frac{\partial\Psi}{\partial\varphi}+\frac{\Omega\,\alpha\,\lambda\,B_{0}}{2}\,r^{2}\,\Psi-\frac{2\,D\,a}{r}\,\Psi+\frac{D\,a^{2}}{r^{2}}\,\Psi+b\,r\,\Psi,
\label{2.2}
\end{eqnarray}

We can go further by taking $\Psi\left(r,\,\varphi,\,z\right)=e^{i\,l\,\varphi+ikz}\,G\left(r\right)$ in Eq. (\ref{2.2}),  however, let us define a new parameter $\varpi$ through the relation:
\begin{eqnarray}
\varpi^{2}=\omega^{2}+4\Omega\,\omega,
\label{2.3}
\end{eqnarray}
where $\omega=\alpha\,\lambda\,B_{0}/m$ is the cyclotron frequency as established in Ref. \cite{lin3}. With this new parameter, let us perform a new change of variables:
\begin{eqnarray}
x=\sqrt{\frac{m\varpi}{2}}\,\,\,\rho,
\label{2.4}
\end{eqnarray}
and thus, we obtain 
\begin{eqnarray}
G''+\frac{1}{x}\,G'-\frac{\tau^{2}}{x^{2}}\,G-x^{2}\,G+\frac{\bar{\mu}}{x}\,G-\bar{\theta}\,x\,G+\bar{\chi}\,G=0,
\label{2.5}
\end{eqnarray}
where $\tau^{2}=l^{2}+2m\,D\,a^{2}$, $\bar{\mu}=\frac{4m\,D\,a}{\sqrt{\frac{m\,\varpi}{2}}}$, $\bar{\theta}=\frac{2m\,b}{\left(\frac{m\,\varpi}{2}\right)^{3/2}}$ and $\bar{\chi}=\frac{2}{m\,\varpi}\left[2m\mathcal{E}-k^{2}+m\,\omega\,l+2m\,l\,\Omega\right]$. 

The behaviour of the function $G\left(x\right)$ at $x\rightarrow\infty$ and $x\rightarrow0$ allows us to write the function $G\left(x\right)$ in terms of an unknown function $\bar{H}\left(x\right)$ as 
\begin{eqnarray}
G\left(x\right)=e^{-\frac{x^{2}}{2}-\frac{\bar{\theta}\,x}{2}}\,x^{\left|\tau\right|}\,\bar{H}\left(x\right),
\label{2.6}
\end{eqnarray}
and then, after substituting Eq. (\ref{2.6}) into Eq. (\ref{2.5}), we obtain
\begin{eqnarray}
\bar{H}''+\left[\frac{2\left|\tau\right|+1}{x}-\bar{\theta}-2x\right]\bar{H}'+\left[\bar{\chi}+\frac{\bar{\theta}^{2}}{4}-2\left|\tau\right|-2+\frac{2\bar{\mu}-\bar{\theta}\left(2\left|\tau\right|+1\right)}{2x}\right]\bar{H}=0,
\label{2.7}
\end{eqnarray}
which is also the biconfluent Heun equation \cite{heun} and $\bar{H}\left(x\right)=H_{\mathrm{B}}\left(2\left|\tau\right|,\,\bar{\theta},\,\chi+\frac{\bar{\theta}^{2}}{4},\,-2\bar{\mu};\,x\right)$ is the biconfluent Heun function.

By following the steps from Eq. (\ref{1.10}) to Eq. (\ref{1.13}), we obtain the relations:
\begin{eqnarray}
b_{1}=\left[\bar{\theta}-\frac{2\bar{\mu}}{2\left|\tau\right|+1}\right]\,b_{0},
\label{2.8}
\end{eqnarray}
and 
\begin{eqnarray}
b_{i+2}=\frac{\bar{\theta}\left(i+1\right)-2\bar{\mu}+\bar{\theta}\left(2\left|\tau\right|+1\right)}{\left(i+2\right)\left(i+2+2\left|\tau\right|\right)}\,b_{i+1}+\frac{2i-\bar{\chi}-\frac{\bar{\theta}^{2}}{4}+2+2\left|\tau\right|}{\left(i+2\right)\left(i+2+2\left|\tau\right|\right)}\,b_{i}.
\label{2.9}
\end{eqnarray}
Besides, in this case we have that the biconfluent Heun series becomes a polynomial of degree $n$ by imposing that \cite{heun}
\begin{eqnarray}
\bar{\chi}+\frac{\bar{\theta}^{2}}{4}-2\left|\tau\right|-2=2n;\,\,\,\,\,\,\,b_{n+1}=0,
\label{2.10}
\end{eqnarray}
for $n=1,2,3,\ldots$. Then, from the condition $\bar{\chi}+\frac{\bar{\theta}^{2}}{4}-2\left|\tau\right|-2=2n$, we obtain 
\begin{eqnarray}
\mathcal{E}_{n,\,l}=\frac{1}{2}\,\sqrt{\omega^{2}+4\Omega\,\omega}\left[n+\left|\tau\right|+1\right]-\frac{1}{2}\,\omega\,l-\frac{2\,\eta^{2}}{m\,\left(\omega^{2}+4\Omega\,\omega\right)}+\frac{k^{2}}{2m}-\Omega\,l,
\label{2.11}
\end{eqnarray}
where $n=1,2,3,\ldots$ and $l=0,\pm1,\pm2,\ldots$. Hence, Eq. (\ref{2.11}) corresponds to the energy levels of the system in a rotating reference frame. In contrast to the analogue of the Landau levels \cite{lin3}, the change in the energy levels and in the degeneracy of them are due to the effects of rotation, the presence of the Kratzer potential and the scalar potential proportional do the radial distance. Furthermore, the effects of rotation give rise to the coupling to the angular momentum quantum number and the angular velocity, which is known as the Page-Werner {\it et al} term \cite{r1,r2,r3}. Note that by taking $\Omega\rightarrow0$, we recover the energy levels (\ref{1.14}).

By following the same analysis of the condition $b_{n+1}=0$ made in Eqs. (\ref{1.15}) and (\ref{1.16}), we obtain that possible values of the cyclotron frequency associated with the lowest energy state ($n=1$) are determined by
\begin{eqnarray}
\left(\omega_{1,\,l}^{2}+4\Omega\,\omega_{1,\,l}\right)^{3/2}&-&\frac{64m\,D^{2}a^{2}}{2\left|\tau\right|+1}\,\left(\omega_{1,\,l}^{2}+4\Omega\,\omega_{1,\,l}\right)+\frac{32\eta\,Da\left(4\left|\tau\right|+3\right)}{2\left|\tau\right|+1}\,\left(\omega_{1,\,l}^{2}+4\Omega\,\omega_{1,\,l}\right)^{1/2}\nonumber\\
&-&\frac{32\,\eta^{2}\left(\left|\tau\right|+1\right)}{m}=0.
\label{2.12}
\end{eqnarray}
Hence, we can see that the possible values of the cyclotron frequency are determined by the quantum numbers of the system $\left\{n,\,l\right\}$ and the parameters that characterizes the rotation, the Kratzer potential and the scalar potential proportional to the radial distance. Again, we can label $\omega=\omega_{n,\,l}$ and rewrite the energy levels (\ref{2.11}) in the form:
\begin{eqnarray}
\mathcal{E}_{n,\,l}=\frac{1}{2}\,\sqrt{\omega^{2}_{n,\,l}+4\Omega\,\omega_{n,\,l}}\left[n+\left|\tau\right|+1\right]-\frac{1}{2}\,l\,\omega_{n,\,l}-\frac{2\,\eta^{2}}{m\,\left(\omega^{2}_{n,\,l}+4\Omega\,\omega_{n,\,l}\right)}+\frac{k^{2}}{2m}-\Omega\,l,
\label{2.13}
\end{eqnarray}
which is the general expression of the spectrum of energy of a neutral particle with an induced electric dipole moment in a rotating reference frame under the influence of a region with a uniform effective magnetic field, the Kratzer potential and a scalar potential proportional to the radial distance.

\section{Conclusions}

We have investigated quantum effects on a neutral particle with no permanent electric dipole moment in both nonrotating and rotating references frames when this quantum particle is subject to scalar potentials. We have also considered a field configuration of electric and magnetic fields that gives rise to a uniform effective magnetic field perpendicular to the $xy$-plane, and then, we have searched for bound states solutions to the Schr\"odinger equation.

In the first case investigated, we have considered a nonrotating reference frame, and thus, obtained the general expression for the energy levels of the system, where we have seen that the presence of the Kratzer potential and the scalar potential proportional to the radial distance modifies the energy levels and breaks the degeneracy of the Landau-type levels \cite{lin3}. However, in search of polynomial solutions to the biconfluent Heun equation, we have seen that there is a restriction on the possible values of the cyclotron frequency characterized by the dependence of the cyclotron frequency on the quantum numbers of the system and the parameters associated with the Kratzer potential and the scalar potential proportional to the radial distance. As an example, we have seen that the possible values of the cyclotron frequency associated with the lowest energy state of the system are determined by a third-degree algebraic equation.

In the second case investigated, it is considered a rotating reference frame, hence, a general expression for the energy levels is obtained, where we have seen that the change in the energy levels and in the degeneracy in contrast to the Landau-type levels \cite{lin3} are due to the effects of rotation and the presence of scalar potentials. Furthermore, we have obtained a contribution to the energy levels given by the coupling to the angular momentum quantum number and the angular velocity, which is called as the Page-Werner {\it et al} term \cite{r1,r2,r3}. Furthermore, in search of polynomial solutions to the biconfluent Heun equation, we have also observed a restriction on the possible values of the cyclotron frequency, where each possible value is determined by the angular velocity of the rotating frame, the parameters associated with the scalar potentials  and the quantum numbers of the system. In particular, we have seen that the possible values of the cyclotron frequency associated with the lowest energy state of the system differs from the case of the nonrotating reference frame and they are determined by Eq. (\ref{2.12}).


\section{Data accessibility statement}

This work does not have any experimental data.

\section{Authors' contributions}

A.B.O. and K.B. conceived the mathematical model, interpreted the results and wrote the paper. A.B.O. made most of the calculations in consultation with K.B. All authors gave final approval for publication.

\section{Competing interests}

We have no competing interests.

\section{Funding Statement}

A.B.O. would like to thank CAPES and K.B. would like to thank CNPq (grant number: 301385/2016-5) for financial support.

\section{Ethics statement}

This research poses no ethical considerations.


\begin{thebibliography}{99}



\bibitem{cross1} G. Wiebusch, J. Main, K. Kr\"uger, H. Rottke, A. Holle and K. H. Welge, {\it Hydrogen atom in crossed magnetic and electric fields}, Phys. Rev. Lett. {\bf62}, 2821 (1989). DOI:10.1103/PhysRevLett.62.2821.

\bibitem{cross2} C. Neumann, R. Ubert, S. Freund, E. Fl\"othmann, B. Sheehy, K. H. Welge, M. R. Haggerty and J. B. Delos, {\it Symmetry Breaking in Crossed Magnetic and Electric Fields}, Phys. Rev. Lett. {\bf78}, 4705 (1997). DOI:10.1103/PhysRevLett.78.4705.

\bibitem{cross3} G. Raithel, M. Fauth and H. Walther, {\it Quasi-Landau resonances in the spectra of rubidium Rydberg atoms in crossed electric and magnetic fields}, Phys. Rev. A {\bf44}, 1898 (1991). DOI:10.1103/PhysRevA.44.1898.

\bibitem{cross4} S. B. Crampton, D. Kleppner and N. F. Ramsey, {\it Hyperfine Separation of Ground-State Atomic Hydrogen}, Phys. Rev. Lett. {\bf11}, 338 (1963). DOI:10.1103/PhysRevLett.11.338

\bibitem{cross5} R. Lutwak, J. Holley, P. P. Chang, S. Paine, D. Kleppner and T. Ducas, {\it Circular states of atomic hydrogen}, Phys. Rev. A {\bf56}, 1443 (1997). DOI:10.1103/PhysRevA.56.1443. 

\bibitem{cross6} O. Dippel, P. Schmelcher and L. S. Cederbaum, {\it Charged anisotropic harmonic oscillator and the hydrogen atom in crossed fields}, Phys. Rev. A {\bf49}, 4415 (1994). DOI:10.1103/PhysRevA.49.4415. 

\bibitem{cross7} G. Raithel, M. Fauth, and H. Walther, {\it Atoms in strong crossed electric and magnetic fields: Evidence for states with large electric-dipole moments}, Phys. Rev. A {\bf47}, 419 (1993). DOI:10.1103/PhysRevA.47.419.

\bibitem{cross8} C. Cuvelliez, D. Baye and M. Vincke, {\it Center-of-mass corrections to the electromagnetic transitions of hydrogen atoms in strong magnetic fields}, Phys. Rev. A {\bf46}, 4055 (1992). DOI:10.1103/PhysRevA.46.4055.

\bibitem{cross9} P. Schmelcher and L. S. Cederbaum, {\it Molecules in strong magnetic fields: Properties of atomic orbitals}, Phys. Rev. A {\bf37}, 672 (1988). DOI:10.1103/PhysRevA.37.672.

\bibitem{cross10} P. Schmelcher and W. Schweizer, {\it atoms and molecules in strong external fields} (Kluwer academic publisher, New York, 1998).

\bibitem{whw} H. Wei, R. Han and X. Wei, {\it Quantum Phase of Induced Dipoles Moving in a Magnetic Field}, Phys. Rev. Lett. {\bf75}, 2071 (1995). DOI:10.1103/PhysRevLett.75.2071.

\bibitem{whw2} H. Wei, X. Wei and R. Han, {\it Wei et al. Reply:}, Phys. Rev. Lett. {\bf77}, 1657 (1996). DOI:10.1103/PhysRevLett.77.1657.

\bibitem{cross11} L. Marmet, H. Held, G. Raithel, J. A. Yeazell and H. Walther, {\it Observation of quasi-Landau wave packets}, Phys. Rev. Lett. {\bf72}, 3779 (1994). DOI:10.1103/PhysRevLett.72.3779. 

\bibitem{cross12} J. A. Yeazell, G. Raithel, L. Marmet, H. Held and H. Walther, {\it Observation of wave packet motion along quasi-Landau orbits}, Phys. Rev. Lett. {\bf70}, 2884 (1993). DOI:10.1103/PhysRevLett.70.2884

\bibitem{lin3} C. Furtado, J. R. Nascimento and L. R. Ribeiro, {\it Landau quantization of neutral particles in an external field}, Phys. Lett. A {\bf358}, 336 (2006). DOI:10.1016/j.physleta.2006.05.069.

\bibitem{landau} L. D. Landau and E. M. Lifshitz, \textit{Quantum Mechanics, the nonrelativistic theory, 3rd Ed.} (Pergamon, Oxford, 1977).

\bibitem{dantas} L. Dantas and C. Furtado, {\it Induced electric dipole in a quantum ring}, Phys. Lett. A {\bf377}, 2926 (2013). DOI:10.1016/j.physleta.2013.09.002

\bibitem{ob} A. B. Oliveira and K. Bakke, {\it On the effects on a Landau-type system for an atom with no permanent electric dipole moment due to a Coulomb-type potential}, Ann. Phys. (NY) {\bf365} 66, (2016). DOI:10.1016/j.aop.2015.12.001

\bibitem{ob2} A. B. Oliveira and K. Bakke, {\it On the Landau system for an atom with no permanent electric dipole moment subject to a linear confining potential}, Int. J. Mod. Phys. A {\bf31}, 1650019 (2016). DOI:10.1142/S0217751X16500196.

\bibitem{dantas1} L. Dantas, C. Furtado and A. L. Silva Netto, {\it Quantum ring in a rotating frame in the presence of a topological defect}, Phys. Lett. A {\bf379}, 11 (2015). DOI:10.1016/j.physleta.2014.10.016.

\bibitem{ob3} A. B. Oliveira and K. Bakke, {\it Effects on a Landau-type system for a neutral particle with no permanent electric dipole moment subject to the
Kratzer potential in a rotating frame}, Proc. R. Soc. A {\bf472}, 20150858 (2016). DOI:10.1098/rspa.2015.0858.

\bibitem{extra} C. A. Dominguez, H. Falomir, M. Ipinza, S. Kohler, M. Loewe and J. C. Rojas, {\it QED vacuum fluctuations and induced electric dipole moment of the neutron}, Phys. Rev. D {\bf80}, 033008 (2009). DOI:10.1103/PhysRevD.80.033008.

\bibitem{extra2} O. Zimmer, C. A. Dominguez, H. Falomir and M. Loewe, {\it Observability of an induced electric dipole moment of the neutron from nonlinear QED}, Phys. Rev. D {\bf85}, 013004 (2012). DOI:10.1103/PhysRevD.85.013004.

\bibitem{kratzer} A. Kratzer, {\it Die ultraroten Rotationsspektren der Halogenwasserstoffe}, Z. Phys. {\bf3}, 289 (1920). DOI:10.1007/BF01327754.

\bibitem{kratzer2} M. R. Setare and E. Karimi, {\it Algebraic approach to the Kratzer potential}, Phys. Scr. {\bf75}, 90 (2007). DOI:10.1088/0031-8949/75/1/015.

\bibitem{kratzer3} G. de A. Marques and V. B. Bezerra, {\it Non-relativistic quantum systems on topological defects spacetimes}, Class. Quantum Grav. {\bf19}, 985 (2002). DOI:10.1088/0264-9381/19/5/310.

\bibitem{kratzer6} D. B. Hayrapetyan, S. M. Amirkhanyan, E. M. Kazaryan and H. A. Sarksiyan, {\it Effect of hydrostatic pressure on diamagnetic susceptibility of hydrogenic donor impurity in core/shell/shell spherical quantum dot with Kratzer confining potential}, Physica E {\bf84}, 367 (2016). DOI:10.1016/j.physe.2016.07.028

\bibitem{kratzer5} D. B. Hayrapetyan, E. M. Kazaryan, L. S. Petrosyan and H. A. Sarkisyan, {\it Core/shell/shell spherical quantum dot with Kratzer confining potential: Impurity states and electrostatic multipoles}, Physica E {\bf66}, 7 (2015). DOI:10.1016/j.physe.2014.09.013.

\bibitem{sag} M. G. Sagnac, {\it L'\'ether lumineux d\'emontr\'e par l'effet du vent relatif d'\'ether dans un interf\'erom\`etre en rotation uniforme}, C. R. Acad. Sci. (Paris) \textbf{157}, 708 (1913).

\bibitem{sag2} M. G. Sagnac, {\it Sur la preuve de la r\'ealit\'e de l'\'ether lumineux par l'exp\'erience de l'interf\'erographe tournant}, C. R. Acad. Sci. (Paris) \textbf{157}, 1410 (1913).

\bibitem{sag5} E. J. Post, {\it Sagnac effect}, Rev. Mod. Phys. \textbf{39}, 475 (1967). DOI:10.1103/RevModPhys.39.475.

\bibitem{r4} B. Mashhoon, {\it Neutron interferometry in a rotating frame of reference}, Phys. Rev. Lett. \textbf{61}, 2639 (1988). DOI:10.1103/PhysRevLett.61.2639.

\bibitem{ac2} Y. Aharonov and G. Carmi, {\it Quantum aspects of the Equivalence Principle}, Found. Phys. {\bf3}, 493 (1973). DOI:10.1007/BF00709117.

\bibitem{cond3} J.-Q. Shen and S.-L. He, {\it Geometric phases of electrons due to spin-rotation coupling in rotating $C_{60}$ molecules}, Phys. Rev. B {\bf68}, 195421 (2003). DOI:10.1103/PhysRevB.68.195421.

\bibitem{cond3a} J. Q. Shen, S. He, and F. Zhuang, {\it Aharonov-Carmi effect and energy shift of valence electrons in rotating C60 molecules}, Eur. Phys. J. D {\bf33}, 35 (2005). DOI:10.1140/epjd/e2005-00027-7.

\bibitem{r13} C.-H. Tsai and D. Neilson, {\it New quantum interference effect in rotating systems}, Phys. Rev. A {\bf37}, 619 (1988). DOI:10.1103/PhysRevA.37.619

\bibitem{anan} J. Anandan and J. Suzuki in {\it Relativity in Rotating Frames, Relativistic Physics in Rotating Reference Frame}, Edited by G. Rizzi and M. L. Ruggiero (Kluwer Academic Publishers, Dordrecht, 2004) p 361-369; arXiv:quant-ph/0305081. 

\bibitem{r14} S.-M. Cui and H.-H. Xu, {\it Berry’s phase in rotating systems}, Phys. Rev. A {\bf44}, 3343 (1991). DOI:10.1103/PhysRevA.44.3343.

\bibitem{r15} S.-M. Cui, {\it Nonadiabatic Berry phase in rotating systems}, Phys. Rev. A {\bf45}, 5255 (1992). DOI:10.1103/PhysRevA.45.5255.

\bibitem{spint1} M. Matsuo, J. Ieda, E. Saitoh and S. Maekawa, {\it Effects of Mechanical Rotation on Spin Currents}, Phys. Rev. Lett. {\bf106}, 076601 (2011). DOI:10.1103/PhysRevLett.106.076601.

\bibitem{spint2} D. Chowdhury, B. Basu, {\it Effect of spin rotation coupling on spin transport}, Ann. Phys. (NY) {\bf339}, 358 (2013). DOI:10.1016/j.aop.2013.09.011

\bibitem{spint3} M. Matsuo, J. Ieda, E. Saitoh and S. Maekawa, {\it Spin-dependent inertial force and spin current in accelerating systems}, Phys. Rev. B {\bf84}, 104410 (2011). DOI:10.1103/PhysRevB.84.104410.

\bibitem{r12} R. Merlin, {\it Rotational anomalies of mesoscopic rings}, Phys. Lett. A {\bf181}, 421 (1993). DOI:10.1016/0375-9601(93)90399-K.

\bibitem{r11} G. Vignale and B. Mashhoon, {\it Persistent current in a rotating mesoscopic ring}, Phys. Lett. A {\bf197}, 444 (1995). DOI:10.1016/0375-9601(94)00981-T.

\bibitem{castro} L. B. Castro, {\it Noninertial effects on the quantum dynamics of scalar bosons}, Eur. Phys. J. C. {\bf76}, 61 (2016). DOI:10.1140/epjc/s10052-016-3904-4.

\bibitem{dkp} M. Hosseinpour and H. Hassanabadi, {\it DKP equation in a rotating frame with magnetic cosmic string background}, Eur. Phys. J. Plus {\bf130}, 236 (2015). DOI: 10.1140/epjp/i2015-15236-8

\bibitem{electro} M. Dvornikov, {\it Galvano-rotational effect induced by electroweak interactions in pulsars}, J. Cos. Astro. Phys. {\bf05}, 037 (2015).

\bibitem{kratzer4} R. K. LeRoy and R. B. Bernstein, {\it Dissociation Energy and Long-Range Potential of Diatomic Molecules from Vibrational Spacings of Higher Levels}J. Chem. Phys. {\bf52}, 3869 (1970). DOI:10.1063/1.1673585.

\bibitem{molecule} S. M. Ikhdair, B. J. Falaye and M. Hamzavi, {\it Nonrelativistic molecular models under external magnetic and AB flux fields}, Ann. Phys. (NY) {\bf353}, 282 (2015). DOI:10.1016/j.aop.2014.11.017.

\bibitem{ct5} I. I. Guseinov and B. A. Mamedov, {\it Evaluation of multicenter one-electron integrals of noninteger u screened Coulomb type potentials and their derivatives over noninteger n Slater orbitals}, J. Chem. Phys. {\bf121}, 1649 (2004). DOI:10.1063/1.1766011.

\bibitem{heun} A. Ronveaux, \textit{Heun's differential equations} (Oxford University Press, Oxford, 1995).

\bibitem{arf} G. B. Arfken and H. J. Weber, {\it Mathematical Methods for Physicists, sixth edition} (Elsevier Academic Press, New York, 2005).

\bibitem{fb4} I. C. Fonseca and K. Bakke, {\it On an atom with a magnetic quadrupole moment subject to harmonic and linear confining potentials}, Proc. R. Soc. A {\bf471}, 20150362 (2015). DOI:10.1098/rspa.2015.0362.

\bibitem{landau4} L. D. Landau and E. M. Lifshitz, \textit{Mechanics, third edition} (Pergamon Press, Oxford, 1980).

\bibitem{landau3} L. D. Landau and E. M. Lifshitz, {\it Statistical Physics - Part 1, 3rd. ed.} (Pergamon Press, New York, 1980).

\bibitem{fb5} I. C. Fonseca and K. Bakke, {\it Rotating effects on the Landau quantization for an atom with a magnetic quadrupole
moment}, J. Chem. Phys. {\bf144}, 014308 (2016). DOI:10.1063/1.4939525.

\bibitem{fb6} I. C. Fonseca and K. Bakke, {\it Rotating effects on an atom with a magnetic quadrupole moment confined to a quantum ring}, Eur. Phys. J. Plus {\bf131}, 67 (2016). DOI 10.1140/epjp/i2016-16067-9.

\bibitem{r1} L. A. Page, {\it Effect of Earth's Rotation in Neutron Interferometry}, Phys. Rev. Lett. \textbf{35}, 543 (1975). DOI:10.1103/PhysRevLett.35.543.

\bibitem{r2} S. A. Werner, J.-L. Staudenmann and R. Colella, {\it Effect of Earth's Rotation on the Quantum Mechanical Phase of the Neutron}, Phys. Rev. Lett. \textbf{42}, 1103 (1979). DOI:10.1103/PhysRevLett.42.1103.

\bibitem{r3} F. W. Hehl and W.-T. Ni, {\it Inertial effects of a Dirac particle}, Phys. Rev. D \textbf{42}, 2045 (1990). DOI:10.1103/PhysRevD.42.2045.





				


\end{thebibliography}
\end{document}